\documentclass[aps,prl,reprint,unsortedaddress]{revtex4-2}

\usepackage{academicons}
\usepackage[colorlinks=true,linkcolor=blue,citecolor=blue,urlcolor=blue]{hyperref}
\usepackage{mhchem} 
\usepackage{float}
\usepackage{graphicx}
\usepackage{multirow}
\usepackage{caption}
\usepackage{makecell}
\usepackage{booktabs}
\usepackage{tabularx}  
\usepackage[normalem]{ulem}
\usepackage{amssymb}
\begin{document}


\title{Probing Vortex $\gamma$ Photons via Nuclear Resonance Fluorescence }





\author{H.\,L.\,Chen}

\affiliation{School of Nuclear Science and Technology, Lanzhou University, Lanzhou 730000, China}
\affiliation{Frontiers Science Center for Rare Isotopes, Lanzhou University,Lanzhou 730000, China}

\author{Y.\,F.\,Niu}

\email{niuyf@sjtu.edu.cn}

\affiliation{School of Physics and Astronomy, Shanghai Jiao Tong University, Key Laboratory for Particle Astrophysics and Cosmology (MoE), Shanghai 200240, China}
\affiliation{Shanghai Key Laboratory for Particle Physics and Cosmology, Shanghai 200240, China}

\author{F.\,Q.\,Chen}

\email{chenfq@lzu.edu.cn}
\affiliation{School of Nuclear Science and Technology, Lanzhou University, Lanzhou 730000, China}
\affiliation{Frontiers Science Center for Rare Isotopes, Lanzhou University,Lanzhou 730000, China}


\begin{abstract}

High-energy vortex $\gamma$ photons offer unique prospects in nuclear physics, astrophysics, and strong-field physics, owing to their distinctive topological structure.  
Yet, their hallmark effects are erased in macroscopic targets, the only practical regime to date, when probed via the total transition probability of photoabsorption. 
Here we show that nuclear resonance fluorescence (NRF) circumvents this limitation.
Using a Bessel-mode description, we demonstrate that for macroscopic targets, the target-averaged angular distribution of scattered photons retains a distinct dependence on the vortex polar angle, which emerges as the sole surviving vortex signature.  
Moreover, by scanning the vortex polar angle instead of the detector angle, we show that NRF can extract the angular momentum of nuclear excited states in a fixed-geometry setup. 
The vortex polar angle, a new degree of freedom in NRF, not only provides a direct quantitative diagnostic for vortex $\gamma$ beams at the MeV energy scale, but also opens a new avenue for exploring orbital angular momentum-induced quantum phenomena in photonuclear physics.

\end{abstract}


\maketitle



Vortex (also called twisted) photons carry a nonzero orbital angular momentum (OAM) together with an intrinsic transverse momentum, which endow them with topological features and spatial symmetries absent in plane-wave states \cite{PhysRevA.45.8185,PhysRevD.83.093001,Knyazev_2018}. 
These unique properties have attracted growing attention for the new insights they bring to light–matter interactions \cite{IVANOV2022103987}. 
Generation techniques now span geometric-phase elements \cite{PhysRevLett.96.163905}, high-harmonic generation \cite{PhysRevLett.113.153901}, helical undulator radiation \cite{PhysRevLett.111.034801}, and laser–plasma interactions \cite{PhysRevLett.117.265001}, making vortex beams routinely available from the optical to the x-ray regime. 
They have thus become a powerful tool in quantum information \cite{mair2001entanglement,fickler2012quantum,erhard2018experimental}, optical communications \cite{wang2012terabit,bozinovic2013terabit,yan2014high}, and atomic physics \cite{schmiegelow2016transfer,PhysRevLett.129.253901,Picon:10}.

In nuclear physics, the application of vortex photons to photonuclear reactions has drawn growing attention in recent years, owing to their unique advantages over conventional probes, although studies so far remain at the theoretical level. 
For instance, when the vortex axis is precisely aligned with the target nucleus, the OAM of the vortex photon can be incorporated into the selection rule, allowing one to manipulate the population of giant resonances of different multipolarities \cite{PhysRevLett.131.202502,guo2025nuclear} and to realize the electric quadrupole excitation of the 8 eV nuclear clock state in $^{229}$Th \cite{PhysRevC.110.064326}. 
In the off-axis case, the controlled weighting of nuclear magnetic substate populations in photon absorption \cite{5md2-ngcf} enables the extraction of the $\gamma$-strength function of the giant quadrupole resonance \cite{XU2024138622} and the detection of chiral structures in condensed-matter systems via vortex Mössbauer spectra \cite{cpl_42_6_061201}. 
Moreover, these distinctive features of nuclear photon absorption can themselves serve as diagnostics for vortex photons \cite{PhysRevLett.131.202502,6m8k-9pcb,Afanasev_2018}.

The practical realization of these studies hinges critically on the energy of the vortex photon, which determines the characteristic transverse scale and thus the required positioning accuracy. 
For the MeV-scale vortex photons considered in most previous photonuclear studies, the target nucleus would need to be placed at a fixed impact parameter $b$ with subpicometer or even femtometer precision, which is far beyond present experimental capabilities. 
In practice, however, nuclei are uniformly distributed over a target area much larger than the characteristic transverse scale of the MeV-scale vortex $\gamma$ beam, which we term a macroscopic target. 
Consequently, the observable effect on such a target must be obtained by averaging over a continuous distribution of impact parameters. 
Remarkably, this averaging completely eliminates all vortex signatures and the associated novel phenomena in the photoabsorption reactions considered, rendering the total transition probability identical to that induced by a plane-wave beam of equal intensity \cite{5md2-ngcf,Maruyama_2024}. 
This implies that, under current experimental conditions with a macroscopic target, no observable difference exists between transition probabilities driven by MeV-scale vortex photons and those driven by plane-wave photons; indeed, one cannot even distinguish whether the incident $\gamma$ beam carries orbital angular momentum or not.

In fact, the detection of high-energy vortex $\gamma$ photons remains a significant challenge. 
Recently, the first experimental evidence for sub-MeV vortex photons, which were generated via all-optical inverse Compton scattering of a Laguerre–Gaussian laser pulse off a laser-wakefield-accelerated electron beam, has been reported \cite{92v4-bzp2}.
However, the identification of the vortex mode relies on the exclusion of classical theoretical explanations. 
Moreover, the disappearance of vortex signatures in photoabsorption invalidates the previously proposed diagnostic based on measuring the total transition probability \cite{PhysRevLett.131.202502,Afanasev_2018}, leaving the detection of high-energy $\gamma$ vortex photons an open problem once again.

It is thus imperative to identify either an alternative photonuclear reaction or a distinct observable beyond the total transition probability that can retain clear vortex signatures of MeV-scale beams even under macroscopic-target conditions. 
Nuclear resonance fluorescence (NRF) \cite{metzger1959resonance,KNEISSL1996349}, the resonant absorption and re-emission of $\gamma$ rays by nuclei, offers a natural pathway. 
Although the absorption step shares the same light-matter interaction as the photoabsorption process for vortex beams, the subsequent emission provides an additional observable: the angular distribution of the scattered photons, which may preserve vortex information after impact-parameter averaging. 
With the advent of quasi-monochromatic, polarized $\gamma$-ray beams at laser Compton backscattering facilities \cite{WELLER2009257,ZILGES2022103903,2022SLEGS,Liu:2024eks}, NRF has matured into a routine spectroscopic tool for studying nuclear excitations below the particle-emission threshold, covering phenomena such as the pygmy dipole resonance \cite{PhysRevLett.89.272502,PhysRevLett.104.072501,PhysRevC.78.064314,PhysRevLett.103.032502,m8b9-lbw3}, two-phonon states \cite{ENDERS2000279,HERZBERG199749,PhysRevLett.70.2880,PhysRevC.108.L051301}, and the scissors mode \cite{HEIL198839,PhysRevC.47.1474,PhysRevLett.117.142501,PhysRevLett.134.082502}. 
A key strength of NRF lies in its ability to determine nuclear spins, parities, and transition strengths in a model-independent manner. 
The angular distribution of scattered photons is governed by the multipolarity and parity of the intermediate excited state: the polar-angle dependence reflects the multipolarity, while the azimuthal-angle dependence encodes the parity. 
Given that vortex photons populate the intermediate state differently than plane-wave photons, it is natural to ask whether such differences manifest in the angular distribution of scattered photons in NRF, even for a macroscopic target.

Therefore, in this Letter we investigate whether NRF can serve as a probe for vortex $\gamma$-ray beams. 
To this end, we establish the theoretical framework for vortex NRF for the first time, and apply it to study the transition probability and angular distribution of the NRF process induced by vortex photons, for both a single nucleus and a macroscopic target. 
We examine whether the angular distribution of the scattered photons retains a characteristic vortex signature for a macroscopic target, which would enable quantitative diagnostics for vortex $\gamma$-ray beams, and further investigate whether vortex NRF offers new advantages for nuclear physics studies compared to traditional NRF experiments.


We adopt the unnormalized standard  Bessel-mode description of a vortex photon  \cite{IVANOV2022103987,kazinski2024excitation,Knyazev_2018}, characterized by its energy $\omega$, total angular momentum (TAM) projection $m_\gamma$, polar angle $\theta_p$, and helicity $\tau$  [see Supplemental Material (SM) \cite{SM}].
Firstly, we consider a single nucleus as the target.
To focus on the vortex signature in the NRF process, we consider the ratio of a vortex to plane-wave photon transition probability for a general $0 \to J_e \to J_f$ channel in an even-even nucleus at impact parameter $\boldsymbol{b} = (b, \phi_b)$:
\begin{gather}
    r(\boldsymbol{b}) = \sum\limits_{M_e}   \mathcal{J}_{m_\gamma - M_e}^2  (\varkappa b) d_{M_e, \tau_i}^{(J_e)2} (\theta_p) , \label{eq:1}
\end{gather}
where $\mathcal{J}_n(x)$ is the Bessel function, $d_{m_1, m_2}^{(L)}$ the Wigner-$d$ function, and $J_e$ ($M_e$) the angular momentum (magnetic projection along the beam axis) of the intermediate state.
The corresponding angular distribution of the scattered photons for the NRF process induced by the vortex photons takes the form
\begin{widetext}
\begin{gather}
    W(\theta, \phi ; \boldsymbol{b}) = \frac{  (2L_f+1) \sum\limits_{ M_f \tau_f}     \biggl|   \sum\limits_{M_e}  e^{i M_e(\phi-\phi_b  + \pi/2 ) } \mathcal{J}_{m_\gamma  - M_e}  (\varkappa b) 
    \begin{pmatrix}
        J_f & L_f & J_e \\
        M_f & M_e - M_f & -M_e
    \end{pmatrix}
    d_{M_e - M_f , \tau_f}^{(L_f)} (\theta)  d_{M_e ,\tau_i}^{(J_e)} (\theta_p)  \biggr|^2}
    {2  \sum\limits_{ M_f M_e}       \mathcal{J}_{m_\gamma  - M_e}^2  (\varkappa b) 
    \begin{pmatrix}
        J_f & L_f & J_e \\
        M_f & M_e - M_f & -M_e
    \end{pmatrix}^2
    d_{M_e ,\tau_i}^{(J_e)2} (\theta_p)   }. \label{eq:2}
\end{gather}
\end{widetext}
In the decay process, we only consider the transition with the lowest multipolarity $L_f = \max \bigl[|J_e - J_f| , 1 \bigr]$, justified by the rapid suppression of higher multipoles in the Weisskopf hierarchy \cite{PhysRev.83.1073}. Then we consider the macroscopic target, where an integration of the transition probability for a single nucleus over impact parameter $\boldsymbol{b}$ is needed. 
Using the longitudinal flux density of vortex beams, we obtain the resulting averaged cross-section for a uniform disk as 
\begin{gather}
    \bar{\sigma}_{z}^{\rm (tw)} \doteq \frac{\int P^{\rm (tw)}(b) \, d^2 b}{\int J_z^{\rm (tw)}(b) \, d^2 b } = \frac{\bar{\sigma}^{\rm (pl)}}{\cos\theta_p}   . \label{eq:3}
\end{gather} 
where $P^{(\text{tw})}(b)$ and $J_z^{(\text{tw})}(b)$ are the transition rate and the longitudinal flux density, respectively.  This expression is in agreement with the photoabsorption case \cite{PhysRevA.92.012705,Afanasev_2018,T19-2025-0054}. 
Here, the $1/\cos\theta_p$ factor in Eq.~\eqref{eq:3} has a transparent origin: each plane-wave component of the Bessel beam propagates at angle $\theta_p$ to the optical vortex axis, so its longitudinal Poynting flux is reduced by $\cos\theta_p$ relative to a plane wave of the same intensity.
Consequently, vortex and plane-wave beams carrying the same intensity produce identical total counting rates of photonuclear reactions, as also observed in the photoabsorption case \cite{5md2-ngcf,Maruyama_2024}.
Therefore, it is more natural to define a cross-section,  using the beam intensity rather than the longitudinal Poynting vector, as 
\begin{gather}
\begin{split}
    \bar{\sigma}_I^{\rm (tw)} \doteq &  \frac{\int P^{\rm (tw)}(b) \, d^2 b / \pi R^2}{ \int w^{\rm (tw)}(r) \, d^3 r /\omega \cdot c / V}  \\
    = &\frac{\int P^{\rm (tw)}(b) \, d^2 b}{\int J_z^{\rm (tw)}(b) \, d^2 b / \cos \theta_p} = \bar{\sigma}^{\rm (pl)}   ,  \label{eq:4}    
\end{split}
\end{gather}
where $w^{\rm (tw)}(r)$ is the energy density of the photon field, $c$ the speed of light, and $V$ the volume.  
The averaged angular distribution of the scattered photons becomes
\begin{gather}
\begin{split}
    &\bar{W} (\theta  ) =   \frac{2L_f+1}{2} \\
    & \times \frac{ \sum\limits_{ M_f \tau_f M_e}     
    \begin{pmatrix}
        J_f & L_f & J_e \\
        M_f & M_e - M_f & -M_e
    \end{pmatrix}^2  d_{M_e - M_f , \tau_f}^{(L_f)2} (\theta)  d_{M_e , \tau_i}^{(J_e)2} (\theta_p)  }
    { \sum\limits_{ M_f M_e}          
    \begin{pmatrix}
        J_f & L_f & J_e \\
        M_f & M_e - M_f & -M_e
    \end{pmatrix}^2  d_{M_e, \tau_i}^{(J_e)2} (\theta_p)} .
    \label{eq:5}  
\end{split}
\end{gather}
Fixing the scattered photon angle $\theta$, one can also obtain the angular distribution $\bar{W}(\theta_p)$ of the incoming vortex polar angle $\theta_p$ from the above equation. 
Further details of the above derivations and explanations are provided in the Supplemental Material \cite{SM}.


Fig.~\ref{Fig1} presents the ratio of NRF transition probabilities for a single nucleus [vortex versus plane-wave photon, see Eq.~\eqref{eq:1}] and the NRF cross-section ratio averaged over a macroscopic target [see Eq.~\eqref{eq:4}]. 
When the nucleus is located on the optical vortex axis ($b = 0$), the cylindrical symmetry enforces angular momentum conservation along the beam axis, $m_\gamma = M_e$ (since $M_i = 0$ for even-even nuclei), forbidding transitions with $J_e<|m_\gamma|$ [panels (a) and (b)]. 
This selection rule has been predicted for nuclear photoabsorption \cite{PhysRevLett.131.202502} and experimentally confirmed for atomic transitions \cite{schmiegelow2016transfer}. 
For off-axis positions ($b \neq 0$), the Bessel function $\mathcal{J}_{m_\gamma - M_e}(\varkappa b)$ allows $M_e\neq m_\gamma$, destroying the selection rule and populating additional magnetic substates \cite{guo2025nuclear,PhysRevA.88.033841}.  
Consequently, intermediate states with $J_e<|m_\gamma|$ become accessible as shown in panels (c) and (d). 
The vortex beam has a characteristic transverse scale $1/\varkappa$ \cite{mcgloin2005bessel}, which is several hundred femtometers for MeV-scale momenta. 
As seen in panels (c) and (d), at $b=3/\varkappa$ the forbidden transitions are already restored, implying that preserving the on-axis rule requires positioning precision far beyond current capability \cite{5md2-ngcf}. 
It is therefore necessary to consider a macroscopic target [panels (e) and (f)]. 
Following Eq.~\eqref{eq:4}, the averaged cross-section defined by the intensity for vortex photons becomes identical to that of the plane-wave case, in agreement with the photoabsorption case \cite{5md2-ngcf,Maruyama_2024}. 
Thus, for the transition probability, NRF leads to the same conclusion as photoabsorption. 
However, unlike photoabsorption, NRF offers an extra observable: the angular distribution of the scattered photons.

\begin{figure}[H]
     \centering
     \includegraphics[width=0.48\textwidth]{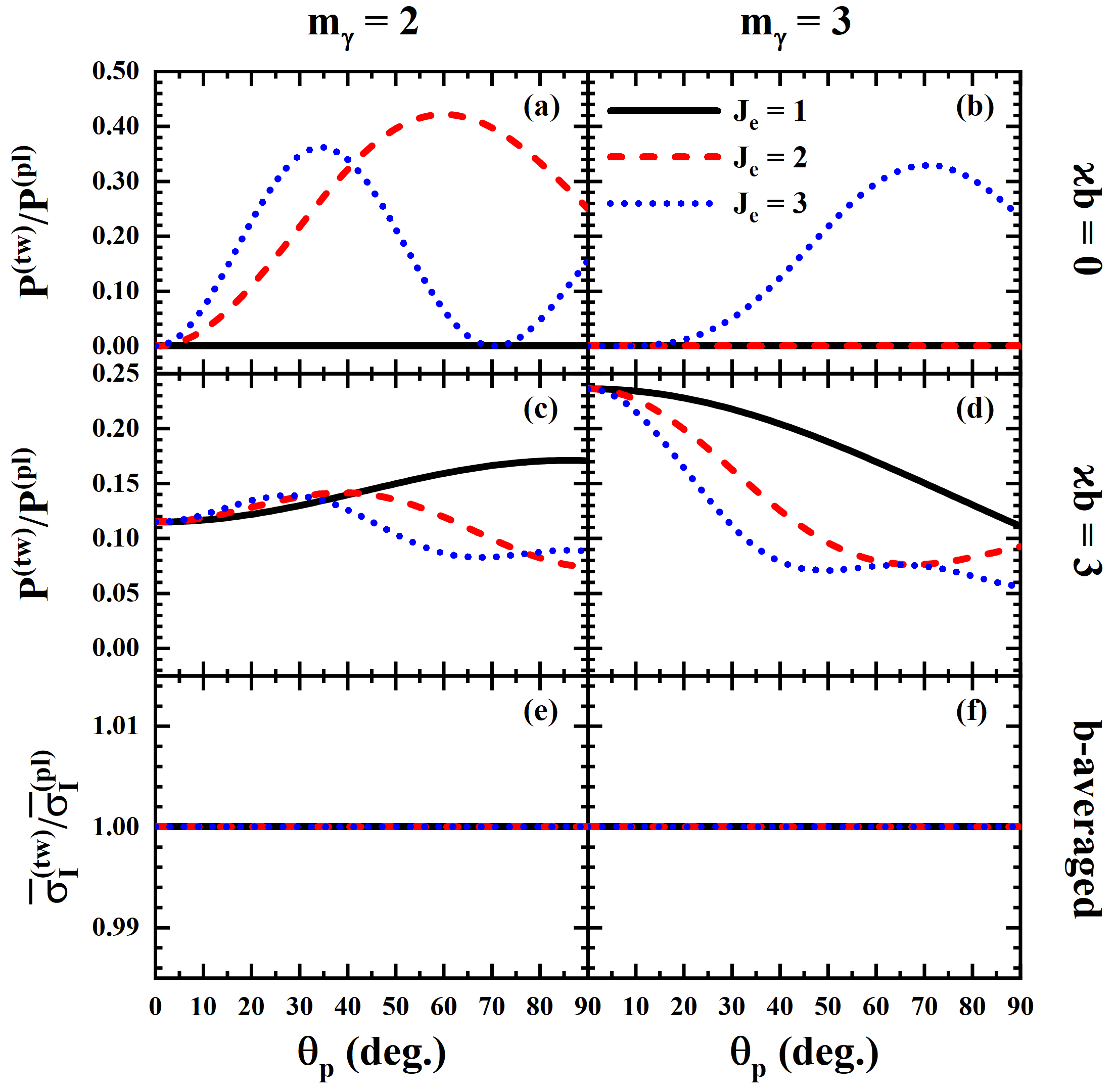}
     \caption{Ratio of NRF transition probabilities for a single nucleus at impact parameter $b$ [panels (a-d)] and ratio of NRF cross-sections $\bar{\sigma}_I$ for a macroscopic target [panels (e) and (f)], plotted as functions of $\theta_p$ for vortex photons with TAM projection $m_\gamma$ and helicity $\tau_i=1$, relative to plane-wave photons. Results for intermediate-state angular momenta $J_e=1,2,3$ are shown as black solid, red dashed, and blue dotted lines, respectively.}
     \label{Fig1} 
\end{figure}


\begin{figure}[H]
     \centering
     \includegraphics[width=0.48\textwidth]{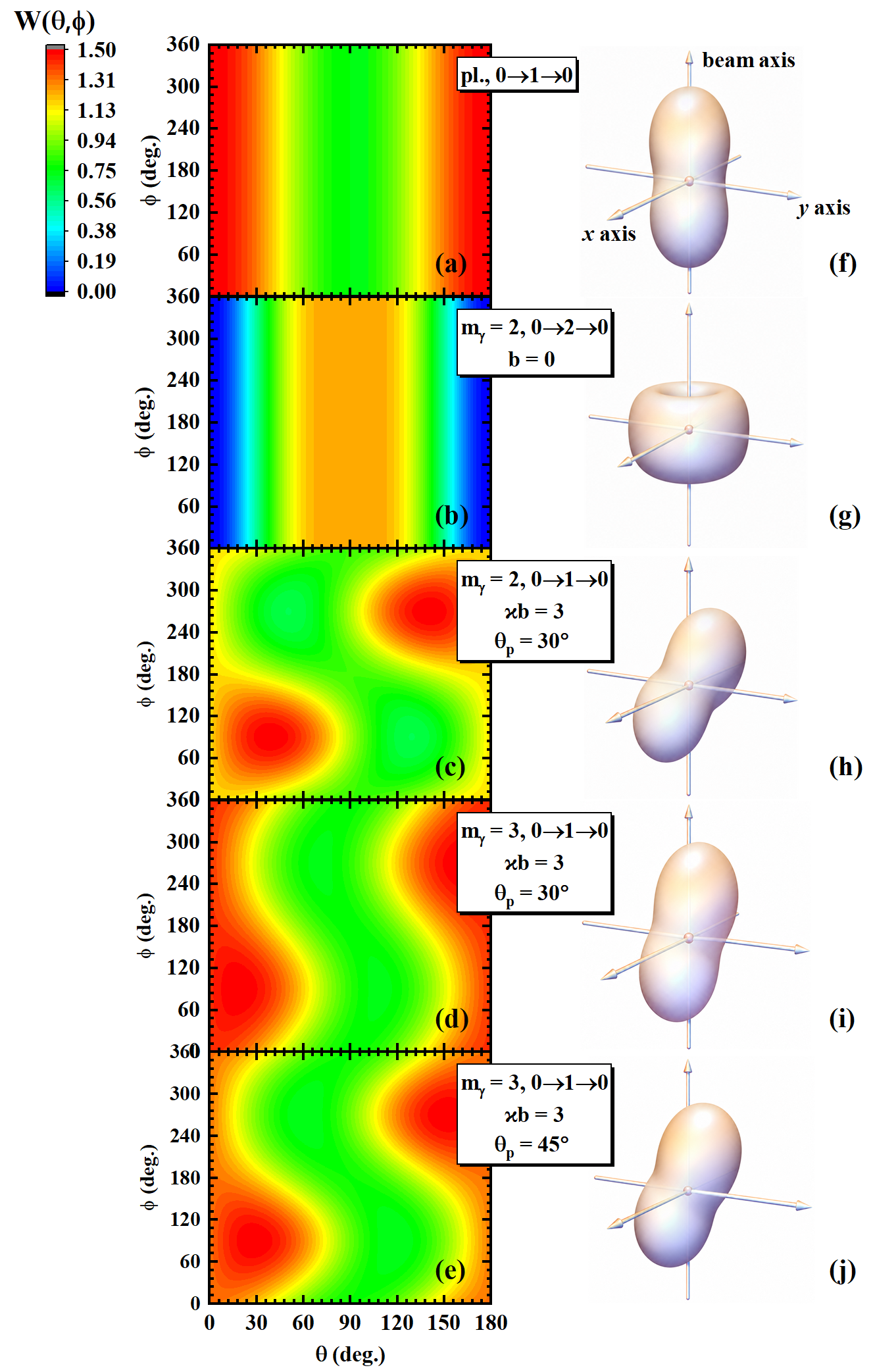}
     \caption{Angular distributions $W(\theta,\phi)$ of scattered photons in NRF on a single-nucleus target for the channels $(J_i \to J_e \to J_f)$ with the highest transition probability: incident plane-wave photons with the channel $0 \to 1 \to 0$ [panels (a) and (f)], incident vortex photons of $m_\gamma = 2$ at $b = 0$ with the channel $0 \to 2 \to 0$ [panels (b) and (g)], incident vortex photons of $m_\gamma = 2$, $\theta_p=30^\circ$ at $\varkappa b = 3$ with the channel $0 \to 1 \to 0$ [panels (c) and (h)], incident vortex photons of $m_\gamma = 3$, $\theta_p=30^\circ$ at $\varkappa b = 3$ with the channel $0 \to 1 \to 0$ [panels (d) and (i)], and incident vortex photons of $m_\gamma = 3$, $\theta_p=45^\circ$ at $\varkappa b = 3$ with the channel $0 \to 1 \to 0$ [panels (e) and (j)]. Panels (a)-(e) are color maps of $W(\theta,\phi)$, while panels (f)-(j) are the corresponding three-dimensional representations with radial distance encoding $W(\theta,\phi)$. The incident photon is circularly polarized with helicity $\tau_i = 1$.}
     \label{Fig2} 
\end{figure}

Therefore, in Fig.~\ref{Fig2} we present the NRF angular distribution for a single nucleus induced by vortex photons for the channels with the highest transition probability, and the plane-wave NRF result is included as a reference. 
Panels (a) and (f) correspond to the dipole transition $0\to1\to0$ induced by plane-wave photons, which yields a $\phi$-isotropic distribution due to axial symmetry when only the substate $M=\tau_i$ is populated. 
For the vortex photons with TAM projection $m_\gamma = 2$, at $b = 0$ the on-axis selection rule forbids the dipole channel, so the quadrupole $0\to2\to0$ dominates.
As shown in panels (b) and (g), $\phi$-isotropy persists owing to cylindrical symmetry, while the $\theta$ profile is reshaped by the quadrupole character, serving as a signature of $m_\gamma$.
At $\varkappa b = 3$, the on-axis selection rule no longer holds, and hence the dipole transition $0 \to 1 \to 0$ regains dominance. 
The angular distribution then resembles the plane-wave case, but with the twisting along the $\phi$ direction, as shown in panels (c-e) and (h-j).
This twisting arises from the broken axial symmetry off-axis, which populates multiple $M_e$ substates, and their interference produces anisotropy in the azimuthal angle $\phi$.
Each substate contributes with amplitude $\mathcal{J}_{m_\gamma - M_e}(\varkappa b) d_{M_e, \tau_i}^{(J_e)}(\theta_p)$, so that the degree of $\phi$-anisotropy is directly controlled by $\theta_p$ and $m_\gamma$ at fixed $b$.
Fixing $\theta_p = 30^\circ$, one can see that the twisting degree differs between $m_\gamma = 2$ [panels (c) and (h)] and $m_\gamma = 3$ [panels (d) and (i)].
While for the case of fixing $m_\gamma = 3$ as shown in panels (d, i) and (e, j), the degree of twisting for the incident vortex photon beam at $\theta_p = 45^\circ$ is greater than that at $\theta_p = 30^\circ$.
In contrast, at $b=0$, the angular distribution remains independent of $\theta_p$ due to the axial symmetry, as shown in panels (b) and (g), and consequently carries no information about the vortex polar angle.
Thus, away from the axis,  the angular anisotropy provides direct access to both the TAM projection $m_\gamma$ 
and the polar angle  $\theta_p$  of the vortex beam.

\begin{figure}[H]
     \centering
     \includegraphics[width=0.48\textwidth]{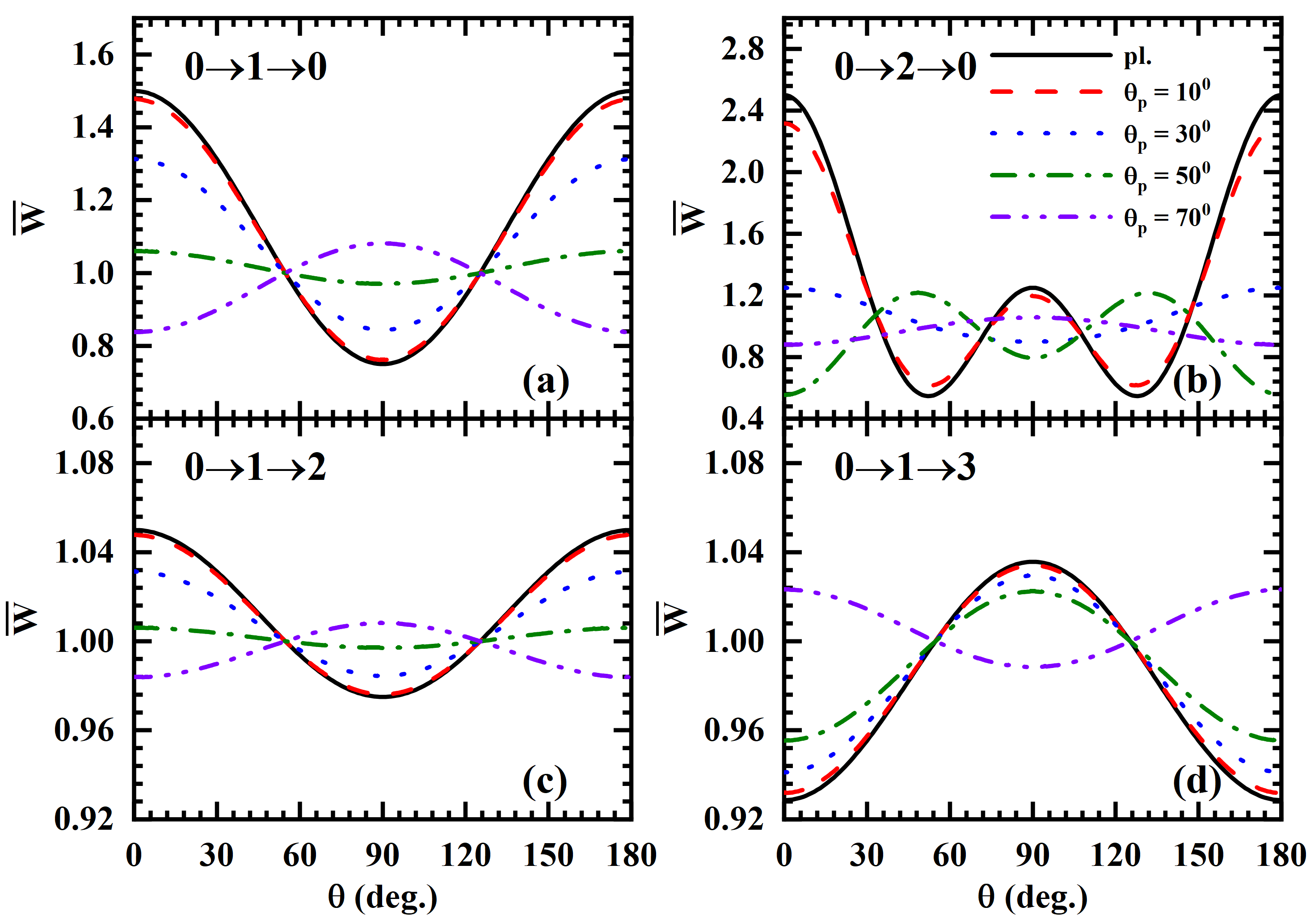}
     \caption{Averaged angular distributions $\bar{W}(\theta)$ [Eq.~\eqref{eq:5}] of scattered photons in NRF on a macroscopic target, for incident plane-wave and vortex photons at several polar angles $\theta_p$. Four transition channels ($J_i \to J_e \to J_f$) are shown: (a) $0 \to 1 \to 0$, (b) $0 \to 2 \to 0$, (c) $0 \to 1 \to 2$, and (d) $0 \to 1 \to 3$. The incident photon is circularly polarized with helicity $\tau_i = 1$.}
     \label{Fig3} 
\end{figure}

The above features, reflecting the vortex signatures, occur only for an ideal single-nucleus target.  
We now turn to the experimentally realistic case of nuclei uniformly distributed over a macroscopic target, for which the averaged angular distribution is given by Eq.~\eqref{eq:5}. 
Fig.~\ref{Fig3} displays $\bar{W}(\theta)$ for four transition channels at several vortex polar angles $\theta_p$. 
Averaging over azimuthal positions restores axial symmetry, so $\bar{W}$ is independent of $\phi$.
This averaging also eliminates all $m_\gamma$-dependent information, leaving the $\theta_p$-dependence as the sole surviving vortex signature. 
Across all channels, increasing $\theta_p$ systematically deforms the angular distribution away from the plane-wave limit, ultimately inverting the positions of maxima and minima.
The $0 \to 1 \to 0$ channel is the most commonly observed NRF transition in even-even nuclei [panel (a)], while the $0\to1\to2$ channel yields a similar shape, since both channels proceed via $L=1$ transitions in excitation and decay, which is a feature that carries over from plane-wave NRF \cite{PhysRevLett.134.022503} to the vortex case. Their cross-sections differ by the branching ratio $\Gamma_2/\Gamma_0$. 
For comparison, the $0\to1\to3$ channel [panel (d)], which has dipole excitation but quadrupole decay, yields a qualitatively different distribution from panels (a) and (c).
The $0 \to 2 \to 0$ channel, in which both excitation and decay are quadrupole, is suppressed by $(kR)^4 \sim 10^{-4}$ \cite{ring1983nuclear,hamilton1974electromagnetic} relative to the dipole channels in the independent-particle model \cite{PhysRev.83.1073}, but exhibits the richest $\theta$ structure and the most pronounced sensitivity to $\theta_p$. 
Thus, $\theta_p$ emerges as a new degree of freedom in vortex NRF, allowing a test of whether the vortex signature is preserved during up-conversion to gamma rays via inverse Compton scattering, as in the setup of Ref. \cite{92v4-bzp2}.

\begin{figure}[H]
     \centering 
     \includegraphics[width=0.48\textwidth]{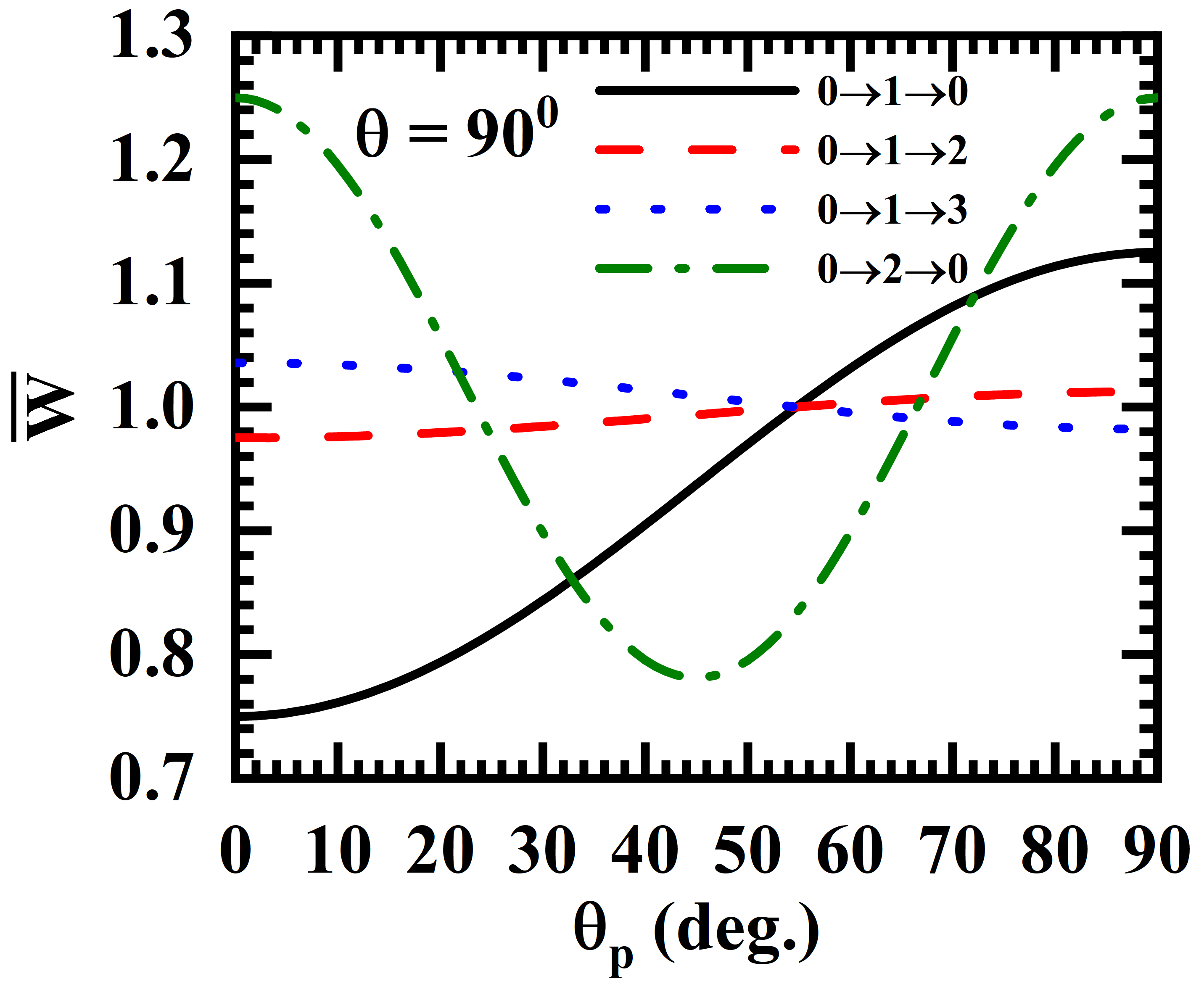}
     \caption{ Averaged angular distribution $\bar{W}(\theta_p)$ of the incoming vortex polar angle $\theta_p$ [Eq.~\eqref{eq:5}] in NRF on a macroscopic target for incident vortex photons, evaluated at the standard NRF observation angle $\theta = 90^\circ$. Four transition channels ($J_i \to J_e \to J_f$) are shown: $0 \to 1 \to 0$, $0 \to 2 \to 0$, $0 \to 1 \to 2$, and $0 \to 1 \to 3$. The incident photon is circularly polarized with helicity $\tau_i = 1$.}
     \label{Fig4} 
\end{figure}

To better illustrate the vortex signature, we fix the polar angle of scattered photons at the standard NRF observation angle $\theta = 90^\circ$ \cite{iliadis2021linear}, which also maximizes the channel separation evident in Fig.~\ref{Fig3}, and show the averaged angular distribution $\bar{W}(\theta_p)$ of the incoming vortex polar angle [Eq.~\eqref{eq:5}] for four channels in 
Fig.~\ref{Fig4}. 
The angular distributions $\bar{W}(\theta_p)$ for $0\to1\to0$ and $0\to2\to0$ depend strongly on $\theta_p$, offering an alternative to conventional $\theta$-scanning: one may fix the detector at a given angle $\theta$ and probe the vortex signature by varying the transverse momentum $\varkappa$ of the beam. 
Moreover, each reaction channel traces a distinct curve as a function of $\theta_p$, enabling identification of the excitation and decay multipolarities from a fixed-angle detector with tunable $\theta_p$. 
We note that with linearly polarized vortex beams \cite{Iorga_2025,PhysRevC.110.064326}, this scheme could in principle also determine the parity of the nuclear excited states, which is a possibility that lies beyond the circularly polarized Bessel-mode description adopted in this work.

In conclusion, we propose NRF as a quantitative, model-independent probe for vortex $\gamma$ photons. 
For the realistic case of a macroscopic target, we demonstrate that the scattered-photon angular distribution retains a distinctive dependence on the vortex polar angle $\theta_p$, the only vortex signature that survives impact-parameter averaging. 
This observable provides a robust diagnostic for vortex $\gamma$  beams. 
Beyond diagnostics, the tunable $\theta_p$ adds a new dimension to NRF spectroscopy, enabling transition-multipolarity assignments at a fixed angle. 
Our results open a new frontier at the interface between vortex-photon physics and nuclear spectroscopy at
$\gamma$-ray facilities, with important implications for both beam diagnostics and photonuclear reaction studies.

\begin{acknowledgments}
This work is supported by 
the Lingchuang Research Project of China National Nuclear Corporation under Grant No. CNNC-LCKY-2024-082, 
the National Natural Science Foundation of China under Grant No. 12075104, No. 12147101, and No. 12447106,
the ``Young Scientist Scheme'' of National Key Research and Development (R\&D) Program under Grant No. 2021YFA1601500,
the Fundamental Research Funds for the Central Universities lzujbky-2023-stlt01,
and the Science and Technology Innovation Leading Talent Project of Gansu Province (Project No. 25RCKA025).
\end{acknowledgments}

\bibliography{references}

\end{document}